\documentclass[10pt]{iopart}
\usepackage{geometry}                
\usepackage[parfill]{parskip}    
\usepackage{graphicx}
\usepackage{amssymb}
\usepackage{epsfig}
\usepackage{epsf}
\usepackage{epstopdf}
\usepackage[hypertex]{hyperref}

\begin{document}
\title{On the distribution of estimators of diffusion constants for Brownian 
motion}

\author{Denis Boyer$^{\bf 1,2}$ and David S. Dean$^{\bf 2}$}

\address{$^{\bf 1}$Instituto de F\'isica, Universidad Nacional Aut\'onoma de M\'exico, D.F. 04510, Mexico\\
$^{\bf 2}$Laboratoire de Physique Th\'eorique -- IRSAMC,
Universit\'e de Toulouse, CNRS, 31062 Toulouse, France}

\ead{\mailto{boyer@fisica.unam.mx}}


\begin{abstract}
We discuss the distribution of various estimators for extracting the diffusion constant of single Brownian
trajectories obtained by fitting the squared displacement of the trajectory. The analysis of the problem can be framed in terms of quadratic functionals of Brownian motion that correspond to the 
Euclidean path integral for simple Harmonic oscillators with time dependent frequencies. Explicit
analytical results are given for the distribution of the diffusion constant estimator in a number of cases
and our results are confirmed by numerical simulations.  

\end{abstract}

\section{Introduction}

\vspace{0.5cm}
The tracking of single particles is a powerful tool to probe physical and 
biological processes at the level of one macromolecule. In particular, the accumulation 
of experimental data in recent years has allowed to test models of diffusive transport 
in cells \cite{saxtonannurev,pederson}. Within aqueous compartments, {\it e.g.} the cell 
cytoplasm, Brownian diffusion is the basic transport mechanism for proteins \cite{dix}. 
Other studies, however, have reported subdiffusive behavior both in membranes 
\cite{saxtonannurev} and in the cytoplasm \cite{cox}, although the microscopic origin 
of anomalous diffusion remains unclear in this context. Crowded environments of the cell 
may cause slower diffusion than in pure water or other solvents, although not necessarily 
subdiffusion \cite{dix}. 

Conflicting results have generated a debate on the methodology for
determining diffusion laws from single particle data, even for simple diffusion
\cite{saxton}. In experiments, trajectories of high temporal and spatial resolution 
are often obtained at the expanse of statistical sample size. Trajectories may be few 
and short due to observation windows limited in space, 
a rapid decay of fluorescent markers or particle denaturation \cite{goulian}.
These limitations complicate the determination of the nature of 
diffusion, {\it i.e.} a precise estimate of the diffusion constant or an 
anomalous exponent. 

In any case, time averaged quantities associated to a trajectory 
may be subjected to large fluctuations among trajectories. 
In the continuous-time random walk model of subdiffusive motion, time-averages of
particle's observables generally are random variables distinct from their ensemble 
averages \cite{barkai}. For instance, the square displacement (after a time lag $t$) 
time-averaged along a given trajectory differs from the ensemble average 
\cite{barkai2}. By analyzing time-average displacements of a particular realization, 
subdiffusive motion can actually look normal, although with strongly differing diffusion 
constants from one trajectory to an other \cite{sokolov}. The Brownian case is 
different, but not as straightforward as often thought. Ergodicity, namely, the equivalence 
of time and ensemble-averages of the square displacement, only holds in this 
case in the infinite sample size limit. In practice, standard fitting procedures applied
to finite (although long) trajectories of a same particle unavoidably 
lead to fluctuating estimates of the diffusion constant. Indeed, variations by orders of 
magnitude have been observed in experiments and simple random walk 
simulations \cite{goulian}. To our knowledge, no analytical results are available on the 
properties of these diffusion constant distributions.

In this article, we present analytical and numerical results on the distributions 
of the diffusion constants estimated from single trajectories. We consider a standard
fitting method based on time-averaged square displacements as well as other
similar procedures amenable to analytical calculations. Generally we show that the problem consists 
of finding the distribution of a quadratic functional of Brownian motion with a time dependent measure. 

The first studies of the quadratic functionals of Brownian motion date back to a classic paper of  Cameron and Martin in 1945 \cite{cam1945}  and the problem has received much interest in the probability community ever since \cite{borodin, don1993, chan1994, rev1999,shi}.
The formulation of path integrals for quantum mechanics provided a powerful tool to analyze
this set of problems using methods more familiar to physicists \cite{fey1965, klein}, here the 
problem appears as the computation of the partition function of a quantum-harmonic oscillator 
with time dependent frequency.  Various quadratic functionals of Brownian motion have 
been intensely studied by physicists \cite{khan1986} using a variety of methods. They arise in a plethora of physical contexts, for polymers in elongational flows \cite{dean1995}, a variety of problems 
related to Casimir/van der Waals interactions and general fluctuation induced interactions \cite{dean2005, pars2006, dean2007, dean2009, dean2010}, where, in harmonic oscillator language, both the frequency and mass depend on time.  Quadratic functionals of Brownian motion also
arise in the theory of electrolytes when one computes the one-loop or fluctuation corrections to the 
mean field Poisson-Boltzmann theory \cite{att1988, rudi1, rudi2, 1dc}. Finally we mention that functionals of Brownian motion also turn out to have applications  in computer science \cite{majumdar}. 

In this paper we use the Feynman-Kac theorem to show that  the generating function, or Laplace transform, of the probability density function of the estimators for diffusion coefficients can be expressed as a solution to an imaginary time Schr\"odinger equation.  This Schr\"odinger equation describes a  particle in a quadratic potential, whose frequency is time dependent. For the choices of time dependent frequency arising in the problem of estimated diffusion constants the resulting Schr\"odinger equation can be solved exactly. The inversion of the resulting Laplace transform to obtain the full distribution cannot be carried out exactly, however we are able to analyze the behavior of the distribution in both the lower and upper tails, thus giving a rather complete analytical description of its behavior.  

In general we find that the main characteristics of the distribution of the estimated diffusion 
coefficient depend little on  the fitting procedure used and in all cases its most probable value is much smaller than  the correct (average) diffusion constant. The probability of measuring a diffusion 
constant lower than average is actually larger than 1/2 (close to 2/3).

\vspace{1cm}

\section{Fits for the diffusion constant of a single trajectory}

\vspace{0.5cm}

Consider a one-dimensional Brownian process $B_t$ of variance 
$\langle B_t^2\rangle=2D_0 t\equiv a_0t$. Without restricting generality,
we set $a_0=1$ and $0\le t\le 1$ in the following. If a particular 
trajectory $B_t$ is available but $a_0$ not known {\it a priori}, 
an estimate $a$ of this parameter can be obtained by performing a fit to 
the diffusion law. Several fitting procedure have been discussed in the 
context of molecule tracking within cells \cite{saxton}. Below, we consider 4 
of them.
 
One of the simplest method consists in calculating a least squares estimate 
based on the minimization of the sum
\begin{equation}\label{F}
F=\int_0^1 [B_t^2-l(t)]^2dt,
\end{equation}
where the diffusion law $l(t)$ can be taken as linear,
\begin{equation}
l(t)=a_Lt, 
\end{equation}
or affine,
\begin{equation}
l(t)=a_At+b_A, 
\end{equation}
typically. Given $B_t$, the minimization of (\ref{F}) with respect
to the constant(s) yields the least squares estimate
\begin{equation}\label{fit1}
a_L=3\int_0^1 tB_t^2dt \quad\quad({\rm FIT 1}) ,
\end{equation}
for the linear fit, and
\begin{eqnarray}\label{fit2}
a_A&=&6\int_0^1 (2t-1)B_t^2 dt\quad\quad({\rm FIT 2}) \label{f21}\\
b_A&=&-2\int_0^1 (3t-2)B_t^2 dt\label{f22}
\end{eqnarray}
for the affine one. 

\begin{figure}
\begin{center}
\epsfig{figure=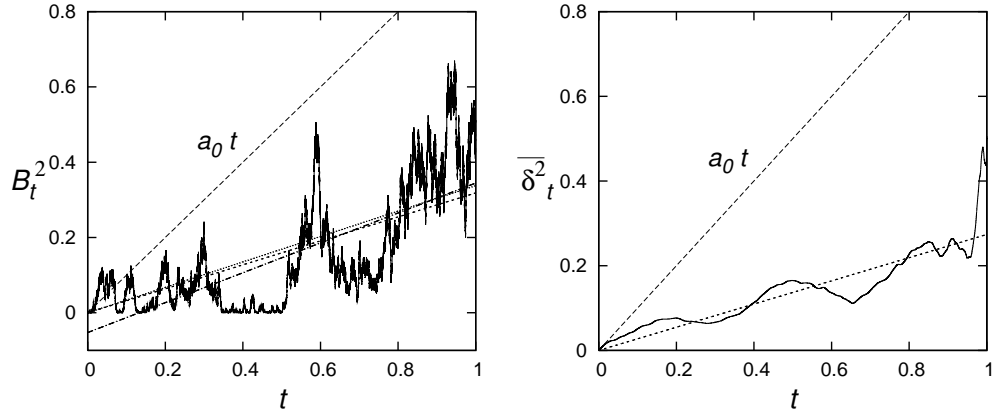,width=2.2in,angle=-90}
\end{center}
\vspace{0.0cm}
\caption{{\bf Left panel:} Square position of a Brownian motion with $a_0=1$ as 
a function of time and the corresponding diffusion laws obtained with the 
fitting methods 1, 2 and 4. For this example, $a_L=0.318$, $a_A=0.397$ and 
$a_{MLE}=0.338$, three values significantly smaller than unity. {\bf Right panel:}
time-average displacement calculated for the same trajectory, where
Fit 3 gives $a_L^{(\delta)}=0.274$. Only at very short times 
$\overline{\delta^2}_t$ follows the ensemble average $a_0t$. 
The trajectory is a random walk of $N=50,000$ steps, with
positions $x_n=\sum_{i=1}^n l_i$ where 
$1\le n\le N$ and $l_i=\pm1$. In the notation of the text, 
$n/N\rightarrow t$ and $x_n^2/N\rightarrow B_t^2$.}
\label{fig:example}
\end{figure}

Another related method, often used in particle tracking experiments
\cite{goulian} and numerical studies \cite{saxton}, consists in least-squares fitting 
the time-averaged square displacement, $\overline{\delta^2}_t$. For a 
finite trajectory, this quantity is defined as
\begin{equation}\label{defdelta}
\overline{\delta^2}_t=
\frac{1}{1-t}\int_0^{1-t}(B_{t+s}-B_{s})^2 ds.
\end{equation}
Due to the ergodicity of normal diffusion processes, at times short compared to 1 
the above average coincides with the ensemble average $\langle B_t^2\rangle$ 
\cite{barkai2}, {\it i.e.}, $\overline{\delta^2}_t\simeq t$ as 
$t\rightarrow 0$. However,
due to practical limitations, experimental trajectories often have
a small number of positions and $\overline{\delta^2}_t$ is analyzed
for all (or a large fraction) of the available intervals $t$, 
like in ref. \cite{goulian}. Similarly, we do not restrict here to $t\ll1$ but 
fit over the whole time 
domain $0\le t\le 1$ instead. As shown by the numerical example of 
Figure (\ref{fig:example}-right) for a random walk with $N=50,000$ 
positions, the expected small $t$ behavior of $\overline{\delta^2}_t$ can 
be restricted to a very small interval compared to the total walk duration. 
Substituting $B_t^2$ by 
$\overline{\delta^2}_t$ in Eq.(\ref{F}) and 
adopting the linear fit, the new estimate simply reads:
\begin{equation}\label{fit3}
a^{(\delta)}_{L}=3\int_0^1 t\ \overline{\delta^2}_t\ dt \quad\quad({\rm FIT 3}).
\end{equation}

Yet another fitting method consists in maximizing the unconditional 
probability of observing the whole trajectory $B_t$, assuming 
that it is drawn from a Brownian process with mean-square displacement 
$at$. Namely, the maximum likelihood estimate (MLE), denoted as 
$a_{MLE}$, is the value of $a$ that maximizes the likelihood of $B_t$,
defined as:
\begin{equation}
L=\prod_{t=0}^1P_a(B_t,t)
=\prod_{t=0}^1 (2\pi at)^{-1/2}\exp\left(-\frac{B_t^2}{2at}\right),
\end{equation}
where $P_a(x,t)$ is the probability density of the Brownian process
with constant $a$. By equating $\partial\ln L/\partial a$ to zero, one obtains
\begin{equation}\label{fit4}
a_{MLE}=\int_0^1dt\ \frac{B_t^2}{t}\quad\quad({\rm FIT 4}).
\end{equation}
The estimates given by the four methods above are 
represented in an example, see Figure (\ref{fig:example}). The numerical values 
are comparable but can differ significantly from unity.

\section{Numerical results}

\begin{figure}
\begin{center}
\epsfig{figure=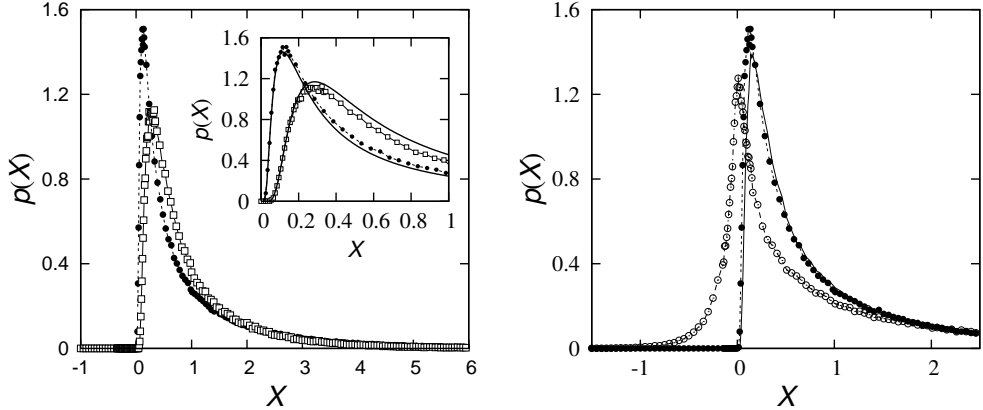,width=2.2in,angle=-90}
\end{center}
\vspace{0.0cm}
\caption{{\bf Left panel:} 
Distributions of the parameters $X=a_L$ ($\bullet$ symbol) and $a_{MLE}$ 
($\Box$ symbol). Inset: zoom of the same plot, where the solid 
lines represent the analytical expressions (\ref{pLan}) and (\ref{pMLEan}) 
valid for small $X$. 
{\bf Right panel:} 
Distributions of the parameters $a_L$ ($\bullet$ symbol) along with 
$a^{(\delta)}_L$ (solid line) and $a_A$ ($\circ$ symbol). Except for 
$a^{(\delta)}_L$, these results are obtained by averaging over $2\ 10^5$ 
random walks with $N=5\ 10^5$ steps.}
\label{fig:distrib}
\end{figure}

The numerical distributions of the random variables $a_L$, $a_{MLE}$, 
$a_L^{(\delta)}$ and $a_A$ are displayed in Figure (\ref{fig:distrib}). 

The distributions are highly asymmetric and peaked near $X=0$, 
far from the average value $\langle X\rangle=1$. The most
probable $X$ is a small positive number in each case, see Table 1.
Although estimates of $X\sim 10$ can be sometimes observed, the median of 
the distribution is lower than $\langle X\rangle$. Namely, the probability
of measuring a diffusion constant lower that the correct value is not 1/2,
but close to 2/3 in all four cases. The probability of measuring a negative $a_A$ 
is not zero in the affine method (as already noticed in ref. \cite{goulian}) 
but close to 0.175. Table 1 summarizes the main properties of the distribution 
functions. 

Importantly, $a_L$ and $a_L^{(\delta)}$ practically obey the same distribution
(Figure (\ref{fig:distrib}-right)), which is somewhat unexpected as $\overline{\delta^2}_t$ 
is a much smoother function than $B^2_t$. Thanks to this similitude, the analytical 
study of the simpler functional (\ref{fit1}), exposed in the next Section, brings many
insights on the behavior of $a_L^{(\delta)}$. Distributions similar to ours for 
$a_L^{(\delta)}$ were determined in ref. \cite{goulian}, both numerically from random walk
simulations and experimentally using R-phycoerythrin proteins in mammalian cells.

\vspace{0.5cm}
\begin{center}
\begin{tabular}{l|c|c|c|c|}
$\ X$ &                    $a_L$   & $a_L^{(\delta)}$ & $a_{MLE}$ & $a_A$  \\
\hline
$\langle X\rangle$         & 1     & 1                & 1         & 1     \\
most probable $X$          & 0.11  & 0.16             & $0.25-0.3$& 0.01  \\
median                     & 0.54  & 0.56             & 0.66      & 0.42  \\
lower 5$\%$                & 0.086 & 0.12             & 0.17      & -0.20 \\
upper 5$\%$                & 3.43  & 3.33             & 2.97      & 4.08  \\
Prob$[X<\langle X\rangle]$ & 0.683 & 0.681            & 0.668     & 0.683 \\
\end{tabular}

\vspace{0.5cm}
Table 1: Main properties of the diffusion constant distributions.
\end{center}
\vspace{1cm}

\section{Feynman-Kac formalism for the generating function }

\vspace{0.5cm}

In general the estimated fit parameters discussed above (FIT1, 2 and 4) are 
quadratic functionals of Brownian motion and take the form
\begin{equation}
X=\int_0^1 B^2_s\ w(s) ds.
\end{equation}
When $w(s)>0$ on $[0,1]$ the quadratic functional is positive and its  generating  function of $X$, is defined  by
\begin{equation}
G(\sigma)=\int_0^\infty p(x)\exp(-\sigma x) dx = 
{\mathbb E}\left[\exp(-\sigma X)\right],
\end{equation}
where $p(x)$ is the probability density function of $X$.
In order to compute $G$ we consider the following average of a quadratic 
functional of Brownian motion:
\begin{equation}
\Psi(x,t) = {\mathbb E}^x\left[\exp(-\sigma\int_t^1 B_s^2 w(s)ds)\right],
\end{equation}
where the expectation above is for a Brownian motion starting at $x$ at 
time $t$. Clearly in this notation we have $G(\sigma)=\Psi(0,0)$. We now 
write  a  Feynman-Kac type formula for $\Psi(x,t)$ by considering how the 
functional evolves in the the time interval $(t,t+dt)$. During this interval
the Brownian motion moves from $x$ to $x+dB_t$, where $dB_t$ is an 
infinitesimal Brownian increment such that $\langle dB_t\rangle=0 $ and 
$\langle dB_t^2\rangle=dt$. Taking into account this evolution
we can write to order $dt$  
\begin{equation}
\Psi(x,t) =\langle {\mathbb E}^{x+dB_t}\left[\exp(-\sigma\int_{t+dt}^1 
B_s^2 w(s)ds)\right](1-dt\sigma w(t)x^2)\rangle
\end{equation}
where the brackets on the right hand side denote the average over $dB_t$. The above may now be written as
\begin{equation}
\Psi(x,t) =\langle \Psi(x+dB_t,t+dt)(1-dt\sigma w(t)x^2)\rangle.
\end{equation}
Expanding to second order in $dB_t$ and $dt$, taking the average over $dB_t$ 
and equating the terms of $O(1)$ and $O(dt)$ we obtain
\begin{equation}
{\partial \Psi\over \partial t}=-{1\over 2}{\partial ^2 \Psi\over \partial
x^2}+\sigma w(t)x^2\Psi, \label{FK}
\end{equation}
which looks like a Schr\"odinger equation in a harmonic, time-dependent 
potential. The boundary condition for this equation is given by $\Psi(x,1)=1$ 
for all $x$. 

It is easy to see that the solution of equation (\ref{FK}) is given by
\begin{equation}
\Psi(x,t)= f(t)\exp(-{1\over 2}g(t)x^2)
\end{equation}
where
\begin{eqnarray}
{d f\over dt }&=& {1\over 2} fg \\
{dg\over dt} &=& g^2 -2\sigma w,
\end{eqnarray}
with the boundary conditions $g(1)=0$ and $f(1)=1$. Now we can eliminate the nonlinearity in the second equation by setting $g=-dh/dt/h$ which gives

\begin{eqnarray}
h{{d f\over dt}}+{1\over 2} f{{dh\over dt}}&=&0 \\
{d^2h\over dt^2} -2\sigma w h&=&0, \label{eqh1}
\end{eqnarray}

with the boundary conditions $h(1)=1$ and $dh/dt(t=1) = 0$. 
In terms of these functions  the Laplace transform is now given by $G(\sigma)=f(0)=1/\sqrt{h(0)}$.
We now make a change of time variable writing
\begin{equation}
{d\tau\over dt} = \sqrt{2 w(t) \sigma},
\end{equation}
assuming for the moment that $w(t)$ is positive. In terms of this new temporal variable equation
(\ref{eqh1}) can now be written as
\begin{equation}
{d^2h\over d\tau^2} +{{d^2\tau\over dt^2}\over \left( {d\tau\over dt}\right)^2}{dh\over d\tau}-h=0.
\label{eqh2}
\end{equation}
In the class of problems we study in this paper (see Eqs.(\ref{fit1}), (\ref{fit2}) and (\ref{fit4})) the form of $w$ is
\begin{equation}
w(t) = (At+C)^{\alpha},
\end{equation} 
with $A$ and $C$ two constants. From this we can choose $\tau$ to be
\begin{equation}
\tau= {\sqrt{8\sigma}\over |A| (\alpha+2)}(At+C)^{\alpha+2\over 2}
\end{equation}
and equation ({\ref{eqh2}) becomes 
\begin{equation}
{d^2h\over d\tau^2} +{\alpha\over (\alpha +2)\tau}{dh\over d\tau}-h=0.
\label{eqh2b}
\end{equation}
The general solution to this equation can be shown to be
\begin{equation}
h(\tau) = \tau^{1\over \alpha+2}\left(DK_{1\over \alpha+2}(\tau)
+EI_{1\over \alpha+2}(\tau)\right),
\end{equation}
where $K_\nu$ and $I_\nu$ are modified Bessel functions \cite{abrom}. The coefficients 
$D$ and $E$ 
are determined from the boundary conditions $h(\tau_1)=1$ and $dh/d\tau=0$ at 
$\tau_1=\tau(1) = \sqrt{8\sigma}(A+C)^{\alpha+2\over 2}/|A|(\alpha+2)$. 
Solving for $D$ and $E$ and using  standard identities for Bessel 
functions \cite{abrom} we find that at 
$\tau_0=\tau(0)=\sqrt{8\sigma}C^{\alpha+2\over 2}/|A|(\alpha+2)$
\begin{equation}\label{h}
h(\tau_0) =\tau_0^{1\over \alpha+2}\tau_1^{\alpha+1\over \alpha+2}
\left(I_{-{\alpha+1\over \alpha+2}}(\tau_1)K_{1\over \alpha +2}(\tau_0)
+K_{-{\alpha+1\over \alpha+2}}(\tau_1)I_{1\over \alpha +2}(\tau_0)\right),
\end{equation}
and thus
\begin{equation}
G(\sigma) = \left[ \tau_0^{1\over \alpha+2}\tau_1^{\alpha+1\over \alpha+2}
\left(I_{-{\alpha+1\over \alpha+2}}(\tau_1)K_{1\over \alpha +2}(\tau_0)
+K_{-{\alpha+1\over \alpha+2}}(\tau_1)I_{1\over \alpha +2}(\tau_0)\right)\right]^{-{1\over 2}}.\label{gen}
\end{equation}

\section{Asymptotic analysis for the probability density function}
The general result equation (\ref{gen})  simplifies in the case where $\tau_0=0$, {\em i.e.} when $C=0$,
which is the case for FIT1 (linear) and FIT4 (MLE). In this case the probability density function of the estimator of the diffusion coefficient $p(x)$ has support on $[0,\infty)$. We start by analyzing the behavior of $p(x)$ at small $x$. 

We proceed by using the small 
argument expansion of $K_\nu$ for $\nu>0$:
\begin{equation}
K_\nu(z)\sim {1\over 2} \Gamma(\nu)({1\over 2}z)^{-\nu} 
\end{equation}
to obtain the exact result
\begin{equation}\label{G1}
G(\sigma)=\left[\Gamma({1\over \alpha + 2})
\left({\sqrt{2\sigma A^\alpha}\over \alpha +2}\right)^{\alpha+1\over \alpha +2} 
I_{-{\alpha+1\over \alpha+2}}
\left({\sqrt{8\sigma A^\alpha}\over \alpha +2}\right)\right]^{-{1\over 2}}.\label{geng}
\end{equation}
The moments of $X$ can then be extracted using the series expansion for 
modified Bessel functions \cite{abrom} which gives
\begin{equation}\label{G2}
G(\sigma)=\left[\Gamma({1\over \alpha + 2})
\sum_{k=0}^\infty {1\over k!}
{\left({2\sigma A^\alpha\over (\alpha +2)^2}\right)^k\over 
\Gamma({1\over \alpha + 2}+k)}\right]^{-{1\over 2}}.
\end{equation}
Without loss of generality we set $A=1$ and find the first two moments 
of $X$ to be given by
\begin{eqnarray}
\langle X\rangle &=& {1\over \alpha+2} \\
\langle X^2 \rangle &=& {3\alpha +7\over (\alpha+2)^2(\alpha+3)}
\end{eqnarray} 
and thus
\begin{equation}
\langle X^2\rangle_c = {2\over (\alpha+2)(\alpha+3)}
\end{equation}
In FIT1 and FIT4, a single estimator for the diffusion constant has the form
\begin{equation}\label{Xalpha}
X_\alpha \equiv (\alpha+2)X=(\alpha+2)\int_0^1 t^\alpha B_t^2dt,
\end{equation}
with $\alpha=1$ and $-1$, respectively, which gives
\begin{equation}
\langle X_\alpha^2\rangle_c = 2(1- {1 \over \alpha+3}),
\end{equation}
From this we see that the MLE estimate of the diffusion coefficient has a 
variance $\langle X_{-1}^2\rangle= 1$ where as the simple linear fit has a 
larger variance $\langle X_{1}^2\rangle= 3/2$. Of course these variances can
be computed directly and the above analysis serves as a check on our formalism
to compute the full probability density function. 

An interesting comparison can be made with the estimator $X_{ep}$ which uses just the 
final value of the mean squared displacement
\begin{equation}
X_{ep} =B_1^2,
\end{equation}
here we find the variance
\begin{equation}
\langle X^2_{ep}\rangle _c =2,
\end{equation}
which is clearly bigger than all the integral estimators above. 
Before embarking on inversion of the generating function $G(\sigma)$ to obtain the 
probability density function $p(x)$, a simple check of our results is to numerically compute
$G(\sigma)$ from our simulation data. In Figure  (\ref{fig:laplace}) 
are shown  the Laplace transforms $G(\sigma)$ obtained from both Eq.(\ref{G1}) 
[or (\ref{G2})] and the numerical distributions $p(x)$, we see that the agreement is perfect.

\begin{figure}
\begin{center}
\epsfig{figure=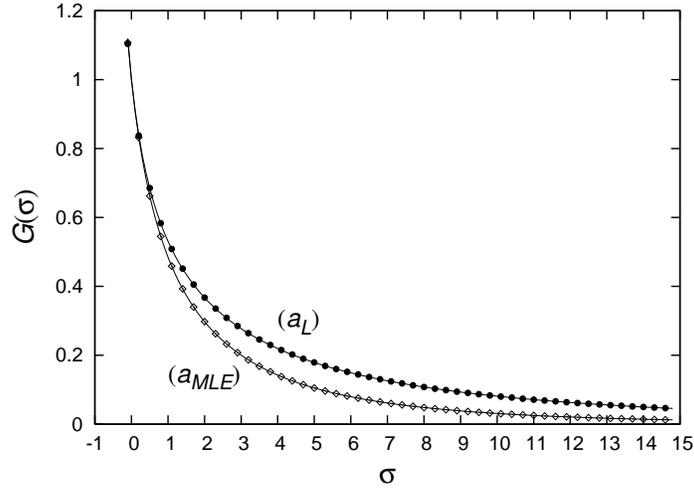,width=2.6in,angle=-90}
\end{center}
\vspace{-0.0cm}
\caption{Laplace transforms of the distributions of $X=a_L$ and
$a_{MLE}$ (cases $\{A=3$, $\alpha=1\}$ and $\{A=1$, $\alpha=-1\}$, 
respectively). The solid lines are given by Eq.(\ref{G1}); the points 
represent the simulations results.}
\label{fig:laplace}
\end{figure}

The behavior of $X$ at small values (when it is always positive) can be 
extracted by examining the characteristic function, or equivalently the 
Laplace transform of the probability density function $p(x)$ of $X$. 
Using the large $z$ asymptotic expansion
\begin{equation}
I_\nu(z)\simeq {1\over \sqrt{2\pi z}}\exp(z)
\end{equation}
and setting $A=1$, we find for large $\sigma$:
\begin{equation}
G(\sigma)\simeq (4\pi)^{1\over 4}\Gamma^{-{1\over 2}}({1\over \alpha + 2})
\left( {2\sigma\over (\alpha +2)^2}\right)^{-{\alpha\over 8(\alpha+2)}}
\exp\left(-{\sqrt{2\sigma}\over (\alpha+2)}\right)
\end{equation}
The behavior of $p(x)$ at small $x$ can now be extracted by noticing 
that the integral
\begin{equation}
I= \int_0^\infty\exp(-\sigma x)\exp(-{d\over x})x^c \ dx
\end{equation}
is dominated by its value at small $x$ and thus can be evaluated by the 
saddle point method as
\begin{equation}
I\simeq \sqrt{\pi\over \sigma} \exp(-2\sqrt{\sigma d})
\left({d\over \sigma}\right)^{2c+1\over 4}
\end{equation}
from which we deduce that for small $x$
\begin{equation}
p(x)\simeq  \pi^{- {1\over 4}}
\Gamma^{-{1\over 2}}({1\over \alpha + 2})
(\alpha+2)^{-{\alpha+4\over 2(\alpha+2)}}\ x^{-{5\alpha+12\over 4(\alpha+2)}}
\exp\left(-{1\over 2 (\alpha+2)^2 x}\right).
\end{equation}
From this  we obtain the probability density of $X=X_{\alpha}$ [Eq.(\ref{Xalpha})] at small $x$ to be:
\begin{equation}\label{px}
p_\alpha (x)\simeq  \pi^{- {1\over 4}}
\Gamma^{-{1\over 2}}({1\over \alpha + 2})
(\alpha+2)^{-{\alpha+4\over 4(\alpha+2)}}\ x^{-{5\alpha+12\over 4(\alpha+2)}}
\exp\left(-{1\over 2 (\alpha+2)x}\right).
\end{equation}
The distribution exhibits an essential singularity at $x=0$, as expected from
the general asymptotic result of Shi \cite{shi}.  
For the linear fit estimate ($\alpha=1$), Eq.(\ref{px}) gives
\begin{equation}\label{pLan}
p_1(x)\simeq c_1\ x^{-{17\over 12}}\exp\left(-{1\over 6x} \right)
\end{equation}
with 
\begin{equation}
c_1=3^{-{5\over 12}}\pi^{- {1\over 4}}\Gamma({1\over 3})^{-{1\over 2}}
\approx 0.29035...,
\end{equation}
and for the MLE ($\alpha=-1$)
\begin{equation}\label{pMLEan}
p_{-1}(x)\simeq c_{-1}\ x^{-{7\over 4}}\exp\left(-{1\over 2x} \right)
\end{equation}
with
\begin{equation}
c_{-1}=\pi^{- {1\over 4}}\approx 0.75112...
\end{equation}
The expressions above compare well with the simulation results at small $x$ 
(Figure (\ref{fig:distrib}-left), inset). The distributions (\ref{pLan}) 
and (\ref{pMLEan}) actually present a maximum at $x^*=2/17\approx 0.118$ and 
$x^{*}=2/7\approx 0.286$, respectively. Despite that the asymptotic results start to 
fail when $x$ becomes too 
large, these values are still in good agreement with the most probable values of Table 1.
A more detailed comparison in the small $x$ regime is displayed in Figure (\ref{fig:scaling}), 
where $p(x)x^{\beta}$ obtained from the numerics is plotted
as a function of $1/x$, with $\beta=17/12$ and $7/4$. The behaviors
at large arguments are nearly indistinguishable from the exponential laws predicted 
by Eqs.(\ref{pLan}) and (\ref{pMLEan}).

\begin{figure}
\begin{center}
\epsfig{figure=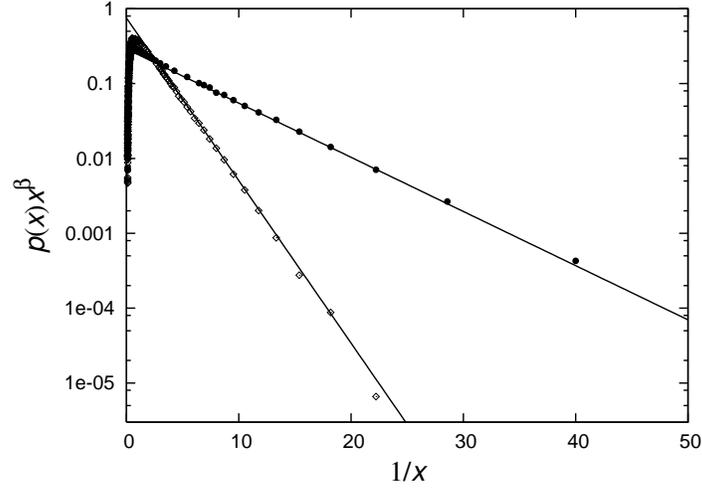,width=2.6in,angle=-90}
\end{center}
\vspace{-0.0cm}
\caption{Rescaled numerical distributions $p(x)x^{\beta}$ with
$\beta=17/12$ (linear fit, black dots) and $\beta=7/4$ (MLE fit, diamonds)
as a function of $1/x$. The solid lines are the analytical
forms $c_1\exp(- {1 \over 6x})$ and $c_{-1}\exp(- {1\over 2x})$ from 
Eqs.(\ref{pLan}) and (\ref{pMLEan}), respectively.}
\label{fig:scaling}
\end{figure}

In order to extract the behavior of the probability distribution for large $x$ we need to examine the 
singularities of the generating function $G(\sigma)$ for $\sigma < 0$, in this regime
\begin{equation}\label{largeg0}
G(\sigma)=\left[
\Gamma(\frac{1}{\alpha+2}) 
\left(\frac{\sqrt{2|\sigma|A^{\alpha}}}{\alpha+2}\right)^{\frac{\alpha+1}{\alpha+2}}
J_{-\frac{\alpha+1}{\alpha+2}}
\left(\frac{\sqrt{8|\sigma|A^{\alpha}}}{\alpha+2}\right)
\right]^{-1/2},
\end{equation}
from the identity $J_{\nu}(z)
=\sum_{k=0}^{\infty}(-1)^{k}(z/2)^{2k+\nu}/[k!\Gamma(k+\nu+1)]$.
This Bessel function of the first kind oscillates and has simple zeros, at these zeros $G$ diverges. 
Let us denote $u^*$ as the lowest 
positive zero of $J_{-\frac{\alpha+1}{\alpha+2}}(u)$. 
When $u\equiv\sqrt{8|\sigma|A^{\alpha}}/(\alpha+2)\rightarrow u^*$ from below,
\begin{equation}\label{largeg1}
\left[J_{-\frac{\alpha+1}{\alpha+2}}(u)\right]^{-1/2}\simeq
\sqrt{
\frac{2|\sigma^*|}{u^*|J_{-\frac{\alpha+1}{\alpha+2}}^{\prime}(u^*)|}
}
(\sigma-\sigma^*)^{-1/2}
\end{equation}
where $\sigma\rightarrow\sigma^*=-u^{*2}(\alpha+2)^2/(8A^{\alpha})$ from above. We now note that
\begin{equation}
\int_0^\infty dx \  {\exp(-\omega x)\over \sqrt{x}} \exp(\sigma x)
= \sqrt{ {\pi\over \omega-\sigma}}\label{largel1}
\end{equation}
for $\omega > \sigma$. 
Comparing Eqs.(\ref{largeg1}) and 
(\ref{largel1}), one deduces from (\ref{largeg0}) the large $x$ behavior:
\begin{equation}
p(x)\simeq
\frac
{2\left(\frac{u^*}{2}\right)^{-\frac{\alpha+1}{2(\alpha+2)}}
|\sigma^*|^{\frac{1}{2}}}
{\sqrt{u^*\Gamma(\frac{1}{\alpha+2})
|J_{-\frac{\alpha+1}{\alpha+2}}^{\prime}(u^*)|} }
\ 
\frac
{e^{-|\sigma^*|x}}
{\sqrt{2\pi x}}. \label{bigx}
\end{equation} 
For the linear fit ($\alpha=1$, $A=3$), one finds $u^*=1.2430...$ and
\begin{equation}\label{largexlinear}
p_{1}(x)\approx 1.1675 \frac{e^{-0.5794x}}{\sqrt{2\pi x}},
\end{equation}
whereas for the MLE ($\alpha=-1$, $A=1$), $u^*=2.4048...$ and
\begin{equation}\label{largexmle}
p_{-1}(x)\approx 1.5212 \frac{e^{-0.7228x}}{\sqrt{2\pi x}}.
\end{equation}
These asymptotic expressions are compared with the numerical results 
in Figure (\ref{fig:largex}-left).

\begin{figure}
\begin{center}
\epsfig{figure=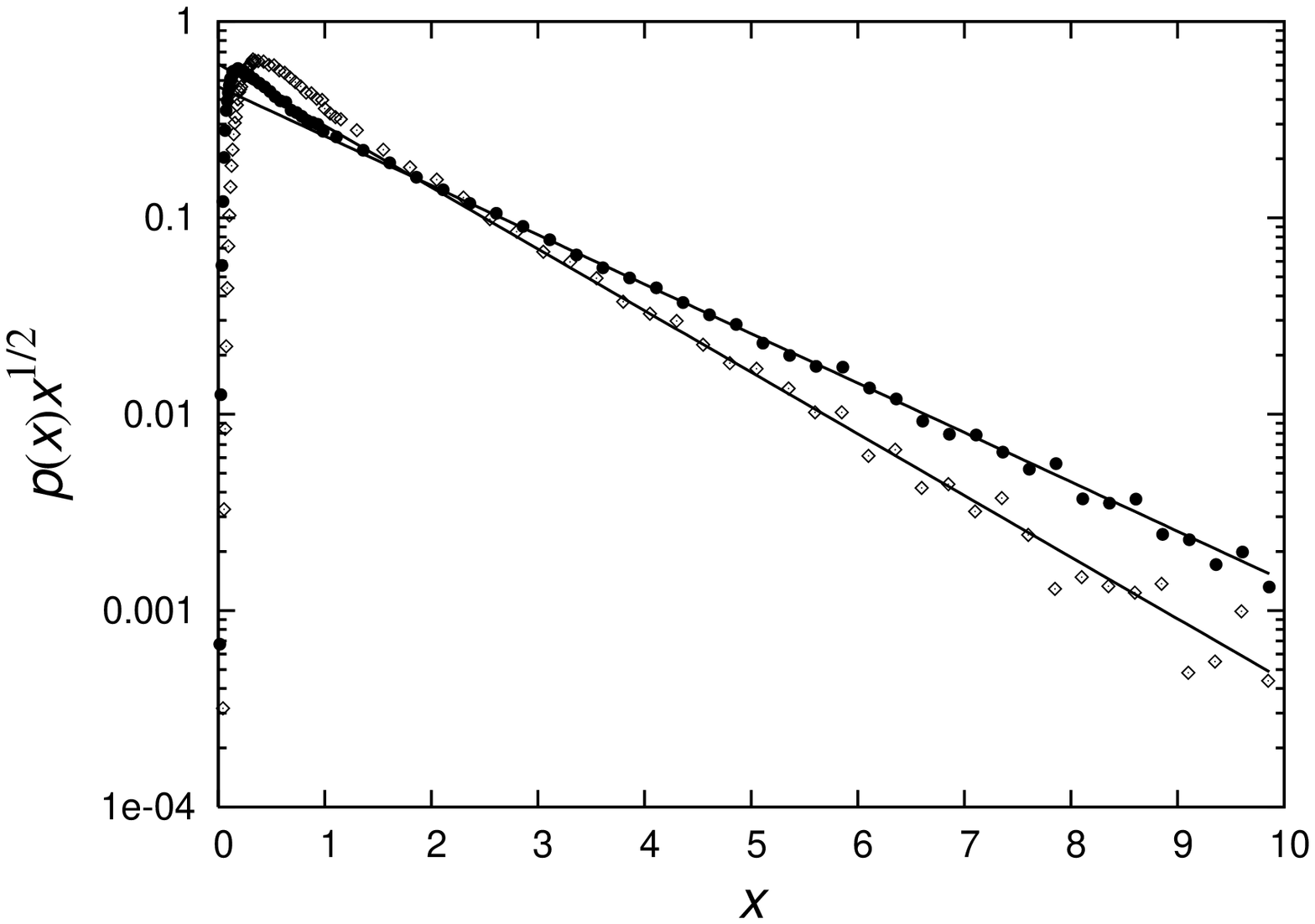,width=2.in,angle=-90}
\epsfig{figure=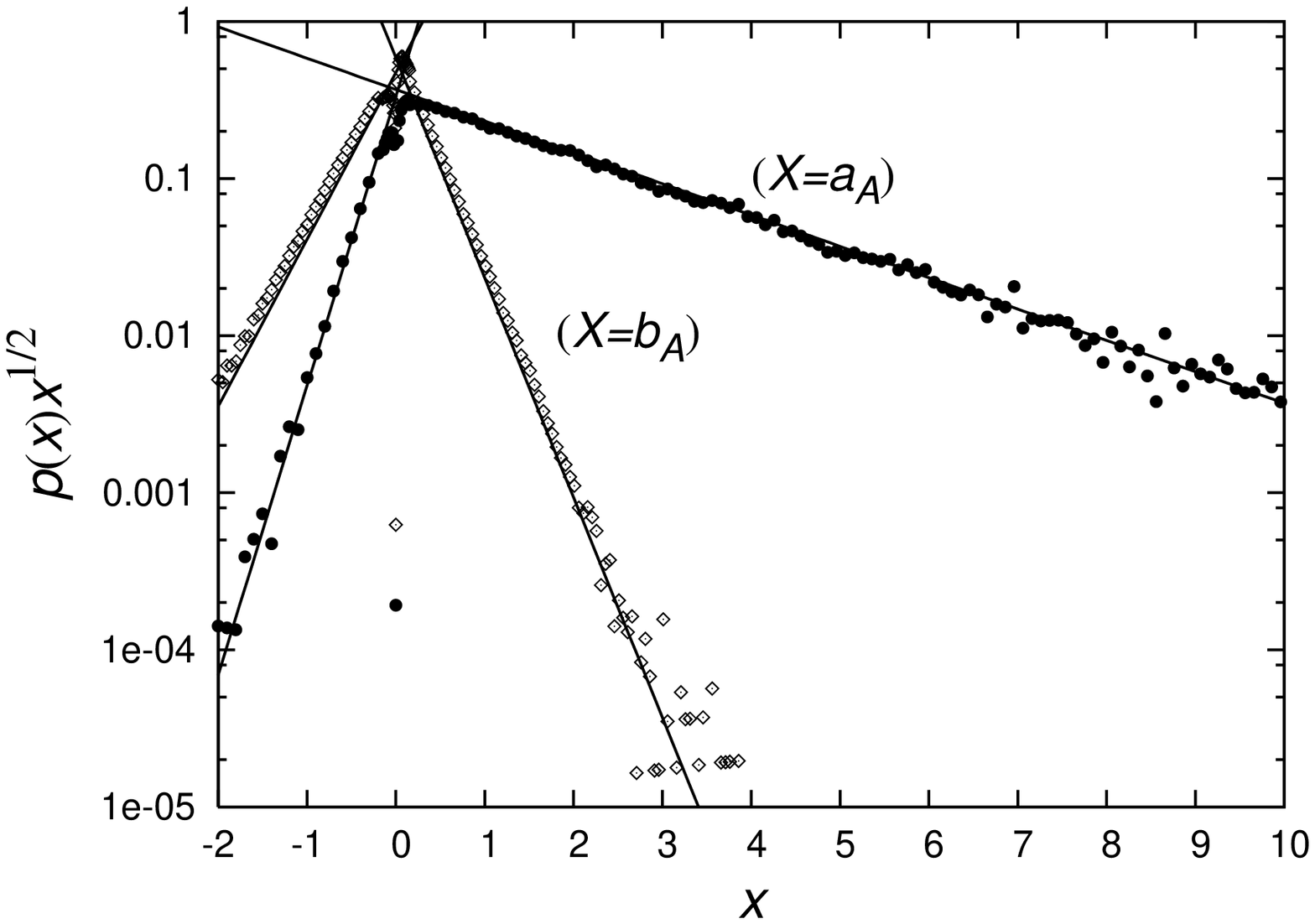,width=2.in,angle=-90}
\end{center}
\vspace{-0.0cm}
\caption{{\bf Left panel:} Rescaled numerical distributions $p(x)x^{1/2}$ for 
the linear (black dots) and MLE (diamonds) fits. The solid lines are the exponential 
laws from Eqs.(\ref{largexlinear}) and (\ref{largexmle}). {\bf Right panel:} Same
quantity for the affine fit (FIT2). The solid lines are the asymptotic forms at 
large $x$ and large $-x$, see Eqs.(\ref{largexfit2}) and (\ref{largemxfit2}).}
\label{fig:largex}
\end{figure}

The interpretation of this result is rather straight forward, if we consider a Gaussian random $Y$ 
variable of mean zero and variance $\gamma^2$ then the probability distribution is
\begin{equation}
p_Y(x) ={1\over \sqrt{2\pi \gamma^2}}\exp(-{x^2\over 2\gamma^2}).
\end{equation}
Now defining $Z=Y^2$ we find that the the probability density function of $Z$ is 
\begin{equation}
p_Z(x) = {\exp(-{x\over 2 \gamma^2})\over \sqrt{2\pi\gamma^2 x}},
\end{equation}
which has the same functional form as equation (\ref{bigx}). This means that for large values of $x$ the
random variable $X$ has the same distribution as a squared Gaussian random variable. This is not 
surprising as the variable $X$ can be viewed as an infinite sum of Gaussian random variables. Note that the full probability density function for the end point estimator $X_{ep}$ is given by (as $\gamma^2=2$)
\begin{equation}
p_{ep}(x) = {\exp(-{0.25 x})\over \sqrt{4\pi x}}, 
\end{equation}
and so the distribution of this simple estimator decays much more slowly that the two integral 
estimators discussed above. 

In the case of the affine fit, FIT2, both the estimators $a_A$ and $b_A$, 
defined in equations (\ref{f21}) and (\ref{f22}), can be  negative as the respective functions $w$ change sign. 
The probability density function is thus two sided.
When $\tau_0$ becomes imaginary in Eq.(\ref{h}), this solution
must be modified by substituting $I_{1/3}(\tau_0)$ and $I_{-1/3}(\tau_0)$ by 
$-J_{1/3}(|\tau_0|)$ and $J_{-1/3}(|\tau_0|)$, respectively \cite{abrom}. In turn, 
when $\tau_1$ becomes imaginary, $I_{2/3}(\tau_1)$ and $I_{-2/3}(\tau_1)$ are
replaced by $J_{2/3}(|\tau_1|)$ and $J_{-2/3}(|\tau_1|)$, respectively.
For large $x>0$ the probability density function can be obtained from the closest zero of $h(\sigma)$ from
zero in the negative direction, denoted by $\sigma^*_-$, and the analysis above goes through
to give
\begin{equation}\label{largexfit2}
p(x)\approx A_-\frac
{e^{-|\sigma^*_-|x}}
{\sqrt{2\pi x}},\quad {\rm with\ } |\sigma^*_-|=0.4596...\ {\rm and}\ 
A_-=0.9239...,
\end{equation}
in the case $X=a_A$. For the variable $X=b_A$, 
one finds $|\sigma^*_-|=3.2229...$ and $A_-=1.4734...$
As $X$ can become negative we also have zeros of $h(\sigma)$ for positive values of $\sigma$; now if the first of these zeros from the origin is $\sigma^*_+$ then the same analysis
as above implies, for $x<0$:
\begin{equation}\label{largemxfit2}
p(x)\approx A_+\frac
{e^{\sigma^*_+x}}{\sqrt{2\pi |x|}},\quad {\rm with\ } \sigma^*_+=4.2439...\ {\rm and}\ 
A_+=0.8381...,
\end{equation}
in the case $X=a_A$. For $X=b_A$, one obtains $\sigma^*_+=2.4485...$\ and $A_+=1.1886...$
These asymptotic results are tested in Figure (\ref{fig:largex}-right) on the two 
sided distribution arising for both the coefficients $a_A$ and $b_A$, showing very good agreement.

\section{Conclusion}

We have shown that a general class of statistical estimators that can be used to extract diffusion constants from the squared displacement of single Brownian trajectories are in fact
quadratic functionals of Brownian motion. Numerically we have seen that such estimators have a 
tendency to yield values which are typically lower than the correct average value. In addition we have
seen that the statistics of the estimated diffusion constants from these trajectories resemble closely
those obtained from fitting the time averaged squared displacement $\overline{\delta^2}_t$, defined in equation (\ref{defdelta}), despite the fact that the resulting trajectory appears much more regular than  an unaveraged Brownian squared displacement, as demonstrated in Figure (\ref{fig:example}-right). An interesting and outstanding problem would be to carry out our analysis for estimators of type
$\delta^2_t$. Such an extension is clearly desirable as it deals with a quantity more commonly used in
single particle tracking experiments. However from a technical point of view the resulting path integrals, while being for quadratic functionals of Gaussian processes, are highly non-local in time and it is 
probable that their evaluation will require the introduction of new mathematical methods. 

Our final analysis was only limited by the problem of carrying out a full Laplace inversion of the 
generating function $G(\sigma)$ to obtain the full probability density function. However we point out that the generating function is actually easy to estimate from numerical data for the purpose of comparison
with our analytical results, as demonstrated in Figure (\ref{fig:laplace}). In addition the generating function
can be inverted analytically in certain asymptotic regimes. When the estimator is always positive, and 
consequently $p(x)=0$ for $x<0$, the behavior of $p(x)$ for small $x$ can be extracted. We find that
it has an essential singularity at $x=0$ and a maximum value, this estimate of the maximum value is
in good agreement for the most likely value of $x$ coming from the full probability density function. 
For positive estimators the large $x$ behavior of $p(x)$ turns out to be that of a squared Gaussian 
random variable, reflecting that fact that the estimator itself is an infinite sum of Gaussian random variables. This remains true when the estimator can have negative values, {\em i.e} when $w(t)$
can change sign. In this case the probability density function for $X$ is that of a Gaussian squared for 
large $x$ as is that of $-X$ for large negative $x$. 

Finally new methods are being introduced into single particle trajectory analysis to estimate diffusion
constants and exponents associated with anomalous diffusion, for instance methods based on 
the mean maximal excursion \cite{mme}, and it would be interesting to examine the distributions
associated with such estimators.
\vskip 1 truecm
\noindent{\bf Acknowledgements:} We would like to than Alain Comtet and Marc M\'ezard for useful
discussions on the subject of this paper. DB would like to thank the Universit\'e de Toulouse (Paul Sabatier) for an invited Professor's position during which this work was initiated.  DSD acknowledges support from the Institut Universitaire de France.


\section*{References}
\begin{thebibliography}{40}

\bibitem{saxtonannurev} 
Saxton M J and Jacobson K 1997
{\it Annu. Rev. Biophys. Biomol. Struct.} {\bf 26} 373--99

\bibitem{pederson}  
Pederson T 2000 {\it Nature Cell Biol.} {\bf 2} E73--74

\bibitem{dix} 
Dix J A and  Verkman A S 2008 {\it Annu. Rev. Biophys.} {\bf 37} 247--63

\bibitem{cox} 
Golding I and  Cox E C 2006 {\it Phys. Rev. Lett.} {\bf 96} 098102

\bibitem{saxton} 
Saxton M J 1997 {\it Biophys. J.} {\bf 72} 1744--53

\bibitem{goulian} 
Goulian M and Simon S M 2000 {\it Biophys. J.} {\bf 79} 2188--98

\bibitem{barkai} 
Rebenshtok A and  Barkai E 2007 {\it Phys. Rev. Lett.} {\bf 99} 210601

\bibitem{barkai2} 
He Y,  Burov S ,  Metzler R and  Barkai E 2008 {\it Phys. Rev. Lett.} {\bf 101} 058101

\bibitem{sokolov} 
Lubelski A,  Sokolov I M  and Klafter J 2008 {\it Phys. Rev. Lett.} {\bf 100} 250602

\bibitem{cam1945} 
Cameron R H  and Martin W T  1945 {\it Bull. Amer. Math. Soc.} {\bf 51} 73--90

\bibitem{borodin}  
Borodin A N 1984 {\it J. Math. Sci.} {\bf 27} 3005--22

\bibitem{don1993} 
Donati-Martin C and Yor M 1993 {\it Adv. Appl. Prob.} {\bf 25} 570--84

\bibitem{chan1994} 
Chan T, Dean D S, Jansons K M  and Rogers L C G 1994 
{\it Comm. Math. Phys.} {\bf 160} 239--57

\bibitem{rev1999} 
Revuz D and Yor M 1999 {\em Continuous Martingales and Brownian Motion}
(Berlin: Springer)

\bibitem{shi} 
Shi Z  1999 {\it Lower tails of quadratic functionals of symmetric stable 
processes},  Preprint Universit\'e Paris 6

\bibitem{fey1965} 
Feynman R P and Hibbs A R 1965 {\em Quantum Mechanics and Path Integrals},
(New York: McGraw-Hill)

\bibitem{klein}
Kleinert H  2006 {\it Path integrals in quantum mechanics, statistics, 
polymer physics and  financial markets} (Singapore: World Scientific)

\bibitem{khan1986}  
Khandekar D C and Lawande S V 1986 {\it Phys. Rep.} {\bf 137} 115--229

\bibitem{dean1995} 
Dean D S and Jansons K M  1995 {\it J. Stat. Phys.} {\bf 79} 265--97

\bibitem{dean2005}
Dean D S  and Horgan R R  2005 {\it J. Phys.: Condens. Matter} {\bf 17} 3473--97

\bibitem{pars2006} 
Parsegian V A 2006 {\em Van der Waals Forces} (Cambridge: Cambridge)

\bibitem{dean2007} 
Dean D S and Horgan R R  2007 {\it Phys. Rev. E. } {\bf 76} 041102

\bibitem{dean2009} 
Dean D S, Horgan R R, Naji A and Podgornik R 2009 {\it Phys. Rev. A} {\bf 79}
040101

\bibitem{dean2010} 
Dean D S, Horgan R R, Naji A and Podgornik R 2010 {\it Phys. Rev. E} {\bf 81}
051117

\bibitem{att1988} 
Attard P, Mitchell J, and Ninham B W  1998 {\it J. Chem. Phys.} {\bf 88} 4987--96

\bibitem{rudi1} 
Podgornik R and Zeks T 1998 {\it J. Chem. Soc. Faraday Trans 2} {\bf 84} 611--31

\bibitem{rudi2} 
Podgornik R  1990  {\it J. Phys. A: Math. Gen.} {\bf 23} 275--84

\bibitem{1dc} 
Dean D S, Horgan R R, Naji A and Podgornik R 2009 {\it J. Chem. Phys.} {\bf 130}
094504

\bibitem{majumdar} 
Majumdar S N  2005 {\it Curr. Sci.} {\bf 89} 2076--92

\bibitem{abrom} 
Abramowitz M  and Stegun I R  1972 {\em Handbook of Mathematical Functions} 
(New York: Dover)

\bibitem{mme} 
Tejedor V, B\'enichou O, Voituriez R, Jungmann R, Simmel F,
Selhuber-Unkel C, Oddershede L B and Metzler R 2010 
{\it Biophys. J.} {\bf 98} 1364--1372

\end{thebibliography}
\end{document}